# Polar solitons in a nonpolar chiral soft matter system

Jia-Hao Chen [1, †], Xing-Zhou Tang [1, †], Yang Ding [1], Susanta Chakraborty [1], Satoshi Aya [3, 4], Bing-Xiang Li [1, 2, *], Yan-Qing Lu [2, *]

[1] College of Electronic and Optical Engineering & College of Flexible Electronics (Future Technology), Nanjing University of Posts and Telecommunications, Nanjing, 210023, China.
[2] National Laboratory of Solid State Microstructures, College of Engineering and Applied Sciences, Nanjing University, Nanjing, 210093, China.
[3] South China Advanced Institute for Soft Matter Science and Technology (AISMST), School of Emergent Soft Matter, South China University of Technology, Guangzhou, 510640, China.
[4] Guangdong Provincial Key Laboratory of Functional and Intelligent Hybrid Materials and Devices, Guangdong Basic Research Center of Excellence for Energy and Information Polymer Materials, South China University of Technology, Guangzhou, 510640, China.
[*] Corresponding authors: bxli@njupt.edu.cn; yqlu@nju.edu.cn.
[†] These authors contributed equally to this work.

## Abstract

Polar solitons, i.e., solitonic waves accompanying asymmetry of geometry or phase, have garnered attention in polar systems, such as ferroelectric or magnetoelectric materials, where they play a critical role in topological transitions and nonreciprocal responses to external fields. A key question is whether such polar solitons can emerge in nonpolar systems, where intrinsic polarity is absent. Here, we demonstrate an unprecedented polar soliton with nematic order in a nonpolar and chiral liquid crystal system by applying an alternating electric field. The soliton is corn-kernel-shaped, displaying a pair of oppositely charged topological defects at its two ends. While head-to-head collision between the solitons leads to repulsion, head-to-tail collision attracts the solitons into a single polar soliton. A rich variety of solitonic kinetics, such as rectilinear translation and circulation motions, can be activated by controlling the voltage and frequency of an electric field. Simulations reveal that the formation of the polar solitons is achieved through balancing the electric and nematic elastic energies, while the flexoelectric effect drives their rotational behaviors. The discovery of polar solitons in nonpolar systems expands the understanding of topological solitons, opening new avenues for dynamic control in soft matter systems, with potential applications in nonreciprocal responsive materials and topological information storage.

# 1 Introduction

Solitons, localized and self-sustaining wave packets that maintain their shape during propagation[1], have become crucial in cosmology, materials science, and biomedicine due to their stability and particle-like characteristics[2-7]. In soft matter systems, solitons appear as optical or mechanical waves with complex structural behaviors[4,8]. Liquid crystals, in particular, offer a compelling platform for soliton research because of their unique combination of ordered molecular alignment and fluidity, which allows the director field to dynamically reorient in response to external fields[9]. This flexibility has enabled the formation of various solitons in liquid crystals, ranging from topological types like skyrmions and hopfions[5,10-13] to non-topological structures such as directrons[14-19], each displaying distinct dynamical properties.

Solitons with intrinsic polarity, recognized as polar solitons, have recently emerged as a promising area of research[6,20-22]. These solitons exhibit dipole-like behavior and respond nonreciprocally to external fields. However, their stable formation in nonpolar environments remains a significant challenge. Nonpolar materials inherently lack the asymmetry required to sustain a polar structure, making it difficult to induce stable polar characteristics in such systems. While polar solitons have been observed in materials with intrinsic polarity, where they find natural support, their demonstration in a nonpolar medium would represent a key advance in understanding solitonic behavior and topological states.

Nematic liquid crystals (NLCs) are typical of nonpolar soft matter, where the alignment of the director field lacks polarity even under an electric field[9]. Despite this non-polarity, various solitons, including skyrmions and directrons, have been successfully created in liquid crystals, revealing rich dynamical properties. The flexoelectric effect offers a pathway to induce polar characteristics in nonpolar liquid crystals by introducing splay and bend deformations aligned with an electric field[23-25]. Stabilizing such induced polarity, however, usually requires careful control of external conditions.

In this work, the generation of a polar soliton within a nonpolar chiral nematic liquid crystal (CNLC) is reported. Through electric annealing and the application of a bias voltage to amplify the flexoelectric effect, a metastable polar soliton with dipole-like interactions during collision and fusion events is achieved. Theoretical modeling and simulations indicate that the soliton's formation, stability, and dynamics result from an interplay among chirality, alternating electric fields, and boundary constraints, with flexoelectric effects providing a directional force under bias. This discovery introduces a new approach to achieving polar order in nonpolar systems, expanding the understanding of controllable soliton interactions in soft matter.

## 2 Results and Discussions

### 2.1 Experimental Configuration and Theoretical Framework

Polar solitons, distinguished by their corn-kernel-like morphology, are generated in a nonpolar chiral system composed of an NLC E7 (97 wt%) and a chiral dopant S811 (3 wt%). The prepared liquid crystal mixture is confined within cells with homeotropic anchoring, having a thickness of $d = 5.0 \pm 1.0$ μm (Supplementary Fig. S1). The obtained equilibrium helicoidal pitch, measured using a Grandjean-Cano cell, is approximately $p \approx 2.9$ μm, Supplementary Fig. S2. An alternating current (AC) electric field $\boldsymbol{E} = (0, 0, E_z)$ is applied perpendicular to the cell substrates, along the $z$-axis (Supplementary Fig. S3). Without a bias voltage, polar solitons remain stationary and stable; however, applying a bias voltage induces circular motion. However, by modulating the applied voltage, the motion of polar solitons can transition between circular and noncircular trajectories. Unique interaction behaviors are observed: head-to-head or head-to-side approaches result in repulsion, while head-to-tail interactions lead to fusion.

To investigate the structural stability and dynamic behavior of polar solitons, we analyzed the system by minimizing the Frank-Oseen free energy density, incorporating contributions from dielectric coupling and flexoelectric effects under the applied electric field. The total free energy density is expressed as follows[17,26-29]:

$$\begin{aligned} f_{\text{CLC}} &= f_{\text{elastic}} + f_{\text{electric}} + f_{\text{flexo}} \\ &= \frac{K_{11}}{2}(\nabla \cdot \mathbf{n})^2 + \frac{K_{22}}{2}(\mathbf{n} \cdot \nabla \times \mathbf{n})^2 + \frac{K_{33}}{2}(\mathbf{n} \times \nabla \times \mathbf{n}) \\ &+ \frac{2\pi K_{22}}{p_0}(\mathbf{n} \times \nabla \times \mathbf{n}) - \frac{\varepsilon_0 \varepsilon_a}{2}(\boldsymbol{E} \cdot \mathbf{n})^2 + \frac{1}{2}\boldsymbol{E} \cdot [e_{11}\mathbf{n} \cdot (\nabla \cdot \mathbf{n}) - e_{33}\mathbf{n} \times \nabla \times \mathbf{n}] \end{aligned} \quad (1)$$

where $\mathbf{n}$ is the director field of the LC, representing the molecular alignment; $K_{11}$, $K_{22}$ and $K_{33}$ are the Frank elastic constants for splay, twist, and bend deformations, respectively; $p_0$ is the equilibrium pitch in the absence of a field; $\varepsilon_0$ is the vacuum permittivity; $\varepsilon_a$ is the dielectric anisotropy; and $e_{11}$, $e_{33}$ are the flexoelectric coefficients.

In this system, chirality introduces an additional twist, expanding the director field to form a fingerprint structure. Meanwhile, the director field aligns with the direction of the applied electric field due to the positive permittivity anisotropy ($\Delta\varepsilon > 0$). The interplay between chirality and the electric field creates a metastable structure recognized as polar solitons. Further, the bias voltage introduces an asymmetry in applied electric field, amplifying the flexoelectric effect and providing an additional force perpendicular to the soliton's motion direction, which drives circular trajectories.

**2.2 Structural Stability and Topological Characteristics of Polar Solitons**

During the observation of optical textures under the polarizing optical microscope (POM), various patterns are observed across a voltage range of 0-15 V and frequencies of 0-10 kHz of the applied voltage, Fig. 1a. Below 2.5 V, the system exhibits a transient translationally invariant configuration (TIC) state as indicated by state I. This TIC state resembles the fingerprint texture but with diffuse stripe boundaries and lasts for a few seconds. The transient nature of this state arises from the chiral properties of CNLCs.

Polar solitons are generated through a specific sequence of voltage applications. Initially, a voltage of 4.3 V with frequency 20 Hz aligns the director field along the $z$-axis, resulting in the homogeneous dark state (state IV). After turning off the field, the system relaxes into the TIC state within 500 ms. Reapplying the electric field of amplitude 2.8 V along the $z$- axis leads to the emergence of polar solitons (state II, Fig. 1b, Supplementary Fig. S4, and Supplementary Movies S1 and S2) and skyrmionic particles[10], including skyrmions and hopfions (state III, Fig. 1c and Supplementary Fig. S5). Fig. 1d and Supplementary Movies S3 illustrate the contraction of the TIC state into the polar solitons. For optimal formation of polar solitons requires a short relaxation duration before reapplying the voltage. Longer relaxation times promote the formation of skyrmionic particles, thereby reducing the prevalence of polar solitons. A detailed analysis of soliton quantities and relaxation times is provided in the appendix (Supplementary Fig. S6).

Numerical modeling applies a normalized voltage scale, with the homeotropic state defined at $\widetilde{U} = 100$. The modeling successfully replicates all experimentally observed patterns. The TIC textures appear below $\widetilde{U} = 70$, while polar solitons emerge within the narrow range of $\widetilde{U} = 70\sim72$. Increasing the voltage within this range reduces the density of polar solitons. Fig. 1e and Supplementary Movies S4 present a simulation of the TIC state relaxing into polar solitons, closely mirroring the experimental process shown in Fig. 1d.

The phase diagram in Fig. 1f summarizes the states observed under various electric field conditions, as well as from simulation results provided in the appendix (Supplementary Fig. S7). While the system behavior is relatively insensitive to frequency, it is strongly influenced by electric field strength. Below 2.5 V, the field cannot overcome elastic forces, and the system relaxes into the fingerprint state. Between 2.5 and 4.0 V, the TIC texture contracts further, forming solitons. In this range, skyrmionic particles may coexist with polar solitons, the latter exhibiting a characteristic corn-kernel shape with a rounded head and tapered tail. These structures display polarization properties, identifying them as polar solitons. However, polar solitons are less stable than skyrmionic particles, which remain stable over a voltage range of approximately 1.2 V, while polar solitons persist over only ~0.1 V. As voltages approaching 4 V, a homeotropic state forms regardless of the frequency, with boundaries tending to align perpendicularly.

## 2.3 Structure of Polar Solitons

The polar soliton exhibits a distinctive corn-kernel-like morphology in three-dimensional space, characterized by a continuous defect ring encircling the structure along its longitudinal $z$- axis (Fig. 2a). This defect ring forms a closed topological structure, classified under the homotopy group $\pi_2(\mathbb{S}^2)$, which represents mappings from the soliton's director field configuration to the spherical order-parameter space. The ring's topological stability is further protected by the homotopy group $\pi_2(\mathbb{S}^2/\mathbb{Z}_2)$, reflecting the nonpolar symmetry of the system, where each director orientation $\mathbf{n}(r)$ is equivalent to $-\mathbf{n}(r)$.

In the $xy$- plane, the defect ring intersects the soliton's head and tail, projecting as $+1/2$ and $-1/2$ topological defects, respectively (Figs. 2b and 2c). These projections arise from symmetry breaking in the two-dimensional plane, where the half-integer winding of the defect ring around the director field induces distinct $+1/2$ and $-1/2$ charges. Between these defects, the director tilts at an angle relative to the $x$- axis, creating a yellow region observable in polarized optical microscopy (POM) images, attributed to birefringence rotation. Under low voltages within one electric field cycle, the defect ring diverges due to chirality, expanding the polar soliton into the TIC state. At higher voltages, the electric field counterbalances chiral forces, causing the defects to attract and annihilate, preventing the reformation of polar solitons when the field strength decreases. This behavior, governed by the interplay of chiral forces and electric fields, occurs within a narrow voltage range and is independent of frequency.

Cross-sections perpendicular to the soliton's *z*- axis (Figs. 2d and 2e) reveal the defect ring appearing as two $-1/2$ defects. Between these defects, the director field exhibits a bilayered arrangement: the outer layer undergoes a clockwise $2\pi$ rotation around the short-axis plane, corresponding to a $+1$ topological charge. Meanwhile, the inner layer aligns more perpendicularly to the plane, resulting in a net topological charge of zero within the cross-section. A detailed representation of polar soliton based on the value of scalar order parameter is illustrated in Supplementary Fig. S8.

Regions between the soliton's ends appear yellow in POM images, while the soliton's boundary along the *z*- axis shows a blue tint. This alignment results in a narrow, dark tip at one end and a wide, bright base at the opposite end, with the base exhibiting an orange hue due to alignment-induced birefringence.

**2.4 Rotational Dynamics of Polar Solitons Under Bias Voltages**

Polar solitons remain stable and exhibit minimal movement when the applied AC field is not accompanied by a bias voltage. However, under a bias voltage, the solitons begin to rotate, as illustrated in Fig. 3a, Supplementary Movie S5. In the experiment, a 2.8 V electric field was applied along the *z*- axis with a 1.1 V bias voltage at a frequency of 100 Hz. Simulations under similar conditions, with $\widetilde{U} = 70$, $\widetilde{U}_b = 12$ and $f = 100$ Hz, reproduced the observed circular motion patterns (Fig. 3b and Supplementary Movie S6). However, the frequency of the applied AC field significantly influences this circular motion. At low frequencies, solitons disappear as the system transitions into a homeotropic state. Conversely, at high frequencies, solitons remain stationary, resembling the condition without a bias voltage. The relationship between bias voltage and the frequency threshold for circular motion exhibits a linear trend, as shown in Fig. 3c.

Referring to the free energy equation (Eq. 1), the system's energy is expressed in terms of splay, twist, and bend, with contributions from chirality and flexoelectricity[30]:

$$\begin{aligned}
f &= \frac{K_{11}}{2}(s)^2 + \frac{K_{22}}{2}(t)^2 + \frac{K_{33}}{2}(\boldsymbol{b})^2 \\
&\quad + \frac{2\pi K_{22}}{p_0}t + \frac{1}{2}\boldsymbol{E}\cdot(e_{11}\mathbf{n}\cdot S - e_{33}\boldsymbol{b}) - \frac{\varepsilon_0\varepsilon_a}{2}(\boldsymbol{E}\cdot\mathbf{n})^2 \\
&= \frac{K_{11}}{2}(s-s_0)^2 + \frac{K_{22}}{2}(t-t_0)^2 + \frac{K_{33}}{2}(\boldsymbol{b}-\boldsymbol{b_0})^2 - \frac{\varepsilon_0\varepsilon_a}{2}(\boldsymbol{E}\cdot\mathbf{n})^2
\end{aligned} \qquad (2)$$

where $s$, $t$, and $\boldsymbol{b}$ are related to splay, twist, and bend. $t_0$ is an additional twist offered by chirality. $s_0$ and $\boldsymbol{b_0}$ are extra splay and bend come from flexoelectricity, representing the effect of flexoelectricity tends to align the director field in the region with larger splay into the direction of the electric field, and adjust the bend to the direction opposite to the electric field.

In the $xy$- plane, the soliton's defect manifests as $+1/2$ and $-1/2$ topological charges. Here the orientation of $+1/2$ defect is defined as the direction of the bend section around, and $-1/2$ defect is described as a triod, each arrow shows the direction of the bend. During one cycle of the electric field, at higher field strengths, the flexoelectric effect causes the direction of $-1/2$ defect tends to align within the $xy$-plane, while the $+1/2$ defect is more likely to align along the $z$- axis. As the voltage decreases, elastic forces realign the $-1/2$ defect along the $z$- axis, while the $+1/2$ defect shifts back to the $xy$- plane. These variations in alignment, driven by differing topological charges, create an orientation bias, establishing distinct head and tail regions.

Without a bias voltage, these alignment shifts average out, maintaining a stable configuration. The light intensity variation of a single soliton over a voltage cycle is shown in the appendix (Supplementary Figs. S9-S11, and Supplementary Movie S7). Applying a bias voltage amplifies the flexoelectric effect, promoting splay and bend along the electric field direction. Due to the asymmetric electric field, the $+1/2$ defect undergoes larger orientation shifts than the $-1/2$ defect, resulting in an average orientation disparity between the head and tail. This orientational difference leads to a lower energy state at the head and a higher energy state at the tail, with the energy difference driving soliton propagation, Fig. 3e.

Fig. 3d shows the force analysis. A stronger electric field during the voltage cycle generates a noticeable electric force, compressing the soliton, while a weaker field allows the chiral effect to expand it. The flexoelectric energy has components associated with both splay and bend. The soliton structure exhibits one side dominated by bend and the other by splay. In region with larger splays, flexoelectricity aligns the director field along the electric field, while in the bend-dominated region, it enhances the bend structure. In the splay-dominated region, the force is aligned with the director field, whereas in the bend-dominated region, it is opposite to the bend direction. Flexoelectric forces on both sides have components along the short axis in the same direction, resulting in a net force perpendicular to the soliton's velocity. These flexoelectric forces generate a transverse force that acts as a centripetal force, driving the circular motion of the polar soliton.

Although the stability of solitons is independent of frequency, it plays a crucial role in governing circular motion dynamics. At low frequencies, the soliton has sufficient time to relax under the forces, and flexoelectric effects more effectively influence circular motion. At high frequencies, the head-tail orientation difference still promotes soliton motion, but the flexoelectric effect has less opportunity to sustain circular motion, requiring a larger bias voltage to enhance flexoelectric forces and support a stable circular motion.

## 2.5 Modulation of Soliton Trajectories

Solitons exhibit either circular propagation or stable positioning depending on the frequency, and the modulation of the bias voltage enables adjustments to their trajectory. A modulated AC field is applied along the *z*- axis, defined by $U(t) = U_m[1 + \frac{1}{2}\sin(2\pi f_m t)]\cos(2\pi f_c t)$, where $U_m$ is the modulation voltage, $f_m$ and $f_c$ represent the modulation and carrier frequencies, respectively. Figs. 4a and 4b and Supplementary Movies S8 and S9 depict the linear propagation of solitons, observed experimentally and reproduced theoretically.

The motion of polar solitons depends on the interplay between modulation and carrier frequencies. Two distinct modes are observed: circular and non-circular, with the latter encompassing both linear and curved trajectories. Fig. 4e provides a phase diagram of soliton motion as a function of frequency. Circular motion is most prominent when the carrier frequency matches the modulation frequency. At higher carrier frequencies in the kilohertz range, solitons deviate from circular trajectories, instead following irregular curved paths.

When $f_m \approx f_c$, the AC field assumes a waveform resembling a square wave with an applied bias voltage, supporting the stable circular motion of polar solitons. This trajectory demonstrates robustness against disturbances. At lower frequencies (Fig. 4e), the voltage required for circular motion is relatively low, allowing solitons to resist larger perturbations and maintain higher velocities, even with notable differences between $f_m$ and $f_c$, as shown in Fig. 4f. However, at frequencies above 100 Hz, small deviations disrupt the centripetal force provided by the electric field, leading to zigzag motion within a cycle and resulting in straight or curved trajectories at the macroscopic scale, Figs. 4c. At frequencies exceeding 1 kHz, increased disturbances prevent solitons from following circular paths, significantly reducing their velocity.

The propagation trajectory of polar solitons can also be altered by abrupt changes in $f_c$, transitioning from circular to linear motion (Fig. 4d and Supplementary Movie S10). For example, with $f_c = 20$ Hz and $f_m = 20$ Hz, the soliton exhibits circular motion. Increasing $f_c$ abruptly from 20 Hz to 60 Hz, while keeping other conditions constant, shifts the trajectory to linear. The initial circular trajectory has a diameter of 10 μm, with a rotation period of 18.4 s and a linear velocity of 3.4 μm/s. Under linear motion, the soliton's velocity reaches 3.3 μm/s.

**2.6 Interaction Behavior of Polar Solitons**

Two topological defects with opposite topological charges at the ends of a polar soliton generate a dipole-like polarization, driving short-range interactions reminiscent of dipole moments. The head of the soliton, characterized by a $+1/2$ defect, serves as the positive pole, while the tail, marked by a $-1/2$ defect, acts as the negative pole. When two solitons approach with their positive poles aligned, mutual repulsion alters their trajectories (Fig. 5a and Supplementary Movies S11 and S12). Analysis of the displacement of solitons "1" and "2" moving horizontally in opposite directions, reveals trajectories along both the *x*- and *y*- axis (Fig. 5b). The *x*- axis displacement shows a linear relationship with time, indicating constant velocity, while the *y*- axis displacement remains stable before reversing direction after the interaction. The dynamic process is detailed in Supplementary Fig. S12.

Polar solitons moving in parallel with aligned directions exhibit mutual attraction. Figs. 5c and 5d and Supplementary Movies S13 and S14 illustrate the fusion process of solitons "3" and "4" moving horizontally. When close enough (approximately 2 μm apart), the head of soliton "3" and the tail of soliton "4" attract and merge into a longer soliton. Due to the influence of the electric field and boundary effects, the length of the merged structure is constrained and relaxes to the size of a single polar soliton, completing the merging process. The resulting soliton, labeled "5" retains the original motion direction of the initial pair. Before merging, the positive component of soliton "3"'s velocity slightly exceeds that of soliton "4". A similar merging behavior is also observed for solitons with aligned negative poles. The detailed dynamic process is shown in Supplementary Figs. S13-S15.

This collision and merging behavior is strongly related to the internal structure of the solitons. As shown in Fig. 5e, two topological defects with opposite charges reside at each end of the polar soliton. When approaching each other with defects of the same sign, the solitons experience a repulsive interaction, resulting in divergence. In contrast, defects with opposite signs attract each other. Such pairs merge in a configuration resembling the fingerprint texture, which remains unstable at a voltage of $\widetilde{U} = 70$. The interaction distance between solitons can be estimated from the energy expression describing defect interactions[31]:

$$F = -\frac{\pi}{2} K \log \frac{L - L_x/2}{2r_c} \qquad (3)$$

where $K$ represents the average elastic constant, $L$ is the distance between two polar solitons, and $r_c$ is the defect core radius and is taken as 1 μm. $L_x$ is the length of the long axis, which is 5 μm. The interaction or repulsion distance can be estimated by reducing the coupling energy by 1% as the soliton propagates from $L$ to $L - r_c$. This leads to the equation $1.01 \log\left(L - \frac{L_x}{2}\right) = \log\left(L - \frac{L_x}{2} - r_c\right) + 0.01 \log(2r_c)$, from which $L$ is estimated as $L \approx 13$ μm.

## 3 Conclusions

This study presents the first demonstration of polar solitons in a nonpolar chiral nematic liquid crystal system, marking a significant advancement in polar soliton research. These solitons, exhibiting dipole-like characteristics, arise from structural rather than electromagnetic propertie. The solitons display distinct, tunable motion patterns—both circular and non-circular—modulated by the carrier and modulation frequencies of the applied electric field. Their interactions, including repulsion between oppositely moving solitons and attraction leading to fusion when moving in the same direction, add a unique aspect to their dynamics. Theoretical modeling and numerical simulations show that the soliton dynamics result from the interplay of chirality, electric fields, and flexoelectric effects. By controlling the electric field waveform and introducing bias voltages, we successfully manipulate soliton motion, demonstrating precise control in nonpolar liquid crystal systems.

These polar solitons, exhibiting robust structures and dynamic motion, hold significant potential for applications in information modulation, encoding, and transmission. Additionally, they enable dynamic control over light propagation and polarization, offering promising applications in nanoelectronics and sensor devices, offering a new paradigm for the creation of electro-optic and magnetic solitons, as well as magnetic monopoles, in nonpolar materials.

## 4 Materials and Methods

**Sample Preparation.** The chiral nematic liquid crystals utilized in this study were prepared by a nematic liquid crystal, E7, which exhibits a positive dielectric anisotropy, with a chiral dopant, S811((S-(+)-2-Octyl 4-(4-hexyloxybenzoyloxy) benzoate). The nematic-isotropic phase transition temperature of E7 is 60.5°C. The pitch, $p$, of the chiral nematic liquid crystals was determined by the concentration, $c$, of the chiral additive with known helical twisting power $h_{\mathrm{HTP}}$, according to the relation $p = 1/(h_{\mathrm{HTP}} \cdot c)$. The CNLC is injected into the cells composed of two glass plates with transparent indium tin oxide (ITO) electrodes at its isotropic state. Two wires are led out from the upper and lower substrates of the liquid crystal cell to provide electric field. The detailed preparation of LC cells and measurement of pitch are described in Supplementary Notes 1-2.

**Simulation Frameworks.** To theoretically simulate our experimental setup, we developed a model using experimentally determined and numerically relaxed TIC states as initial conditions. The free energy functional for the tensorial order parameter (**Q**) is expressed as $\mathbf{Q} = S(\mathbf{nn} - \mathbf{I}/3)$, where **n** represents the director field's unit vector, and $S$ is the scalar order parameter. An AC electric field, negatively coupled to the director through the liquid crystal's anisotropic permittivity, is applied. This setup confines the liquid crystal between two walls, with a circular design pattern featuring reduced anchoring on the upper layer. The evolution of **Q** is governed by the Ginzburg-Landau equation:

$$\frac{\partial \mathbf{Q}}{\partial t} = \Gamma \mathbf{H};$$
$$\mathbf{H} = -\left(\frac{\delta F}{\delta \mathbf{Q}} - \frac{I}{3}\mathrm{Tr}\frac{\delta F}{\delta \mathbf{Q}}\right) \quad (4)$$

Here, $\Gamma$ denotes a collective rotational diffusion constant that dictates the rate of relaxation, and $\mathbf{H}$ represents the molecular field.

The system's free energy is characterized by

$$F = \int [f_{\mathrm{LdG}} + f_{\mathrm{elas}} + f_{\mathrm{flex}} + f_{\mathrm{diel}}] \, dV \quad (5)$$

where the bulk free energy density encompasses contributions from enthalpic ($f_{\mathrm{LdG}}$), elasticity ($f_{\mathrm{elas}}$), flexoelectricity ($f_{\mathrm{flex}}$), and dielectric ($f_{\mathrm{diel}}$) energies, which are representing in terms of director field $\mathbf{n}$ in equation (1) and the tensor forms are discussed in Supplementary Notes 3-4. The strong surface anchoring is taken in this model as well as neglecting the effect from ions and induced current.

The simulation employs specific parameters: A = 10, $U$ = 3.5, $K_{11}$ = 6.1, $K_{22}$ = 2.2, $K_{33}$ = 5.8, $e_{11}$ = 5, $e_{33}$ = −5, $\varepsilon_a$ = 5, $\Gamma$ = 0.1 and $q_0$ = 0.1. These parameters were meticulously chosen based on experimental insights and refined through non-dimensionalization to ensure accurate simulation outcomes.

**Generation of Polar Solitons.** The electric field was applied perpendicularly to the chiral nematic liquid crystal cells at 25°C on a hot stage (HCS402XY, Instec) using a waveform generator (AFG1022, Agilent) and a high voltage amplifier (ATA-2041, Aigtek).

**Microscopic Observations.** Samples and the motions of topological solitons were observed using a polarizing microscope (ECLIPSE Ci-POL). Images and videos were recorded using a digital microscope camera (E3ISPM09000KPB, Kainuo, China) and a high-speed camera (SH3-502, SinceVision, China). During the motion measurements, the frame rate of the camera was adjusted to 20-2000 frames per second. The voltage values are measured using a multimeter (DMM6500, Keithley).

**Data Analysis.** The motion of the solitons is tracked and analyzed by using open-source ImageJ[32] and Fiji[33] software. The position information and number density of the solitons are extracted for each frame through a plugin of ImageJ, TrackMate. The temporal evolution of velocity order is obtained by analyzing the positional information of solitons between frames of movies.

**Acknowledgments**

The work is supported by the National Key Research and Development Program of China (No.2022YFA1405000), the National Natural Science Foundation of China (No. 62375141), the Natural Science Foundation of Jiangsu Province, Major Project (No. BK20243067).

**Author Contributions**

J.H.C. and X.Z.T. contributed equally to this work. J.H.C., and Y.D. performed the experimental studies. X.Z.T. performed the theoretical analysis and numerical simulations. J.H.C., X.Z.T., Y.D., S.C., S.A., B.X.L., and Y.Q.L. analyzed the data and discussed the mechanisms. B.X.L. and Y.Q.L. directed the research. All the authors contributed to the discussion and wrote the manuscript.

**Data Availability**

The data that support the findings of the study are available from the corresponding author upon reasonable request.

**Competing Interests**

The authors declare no competing interests.

**Figures 1-5**

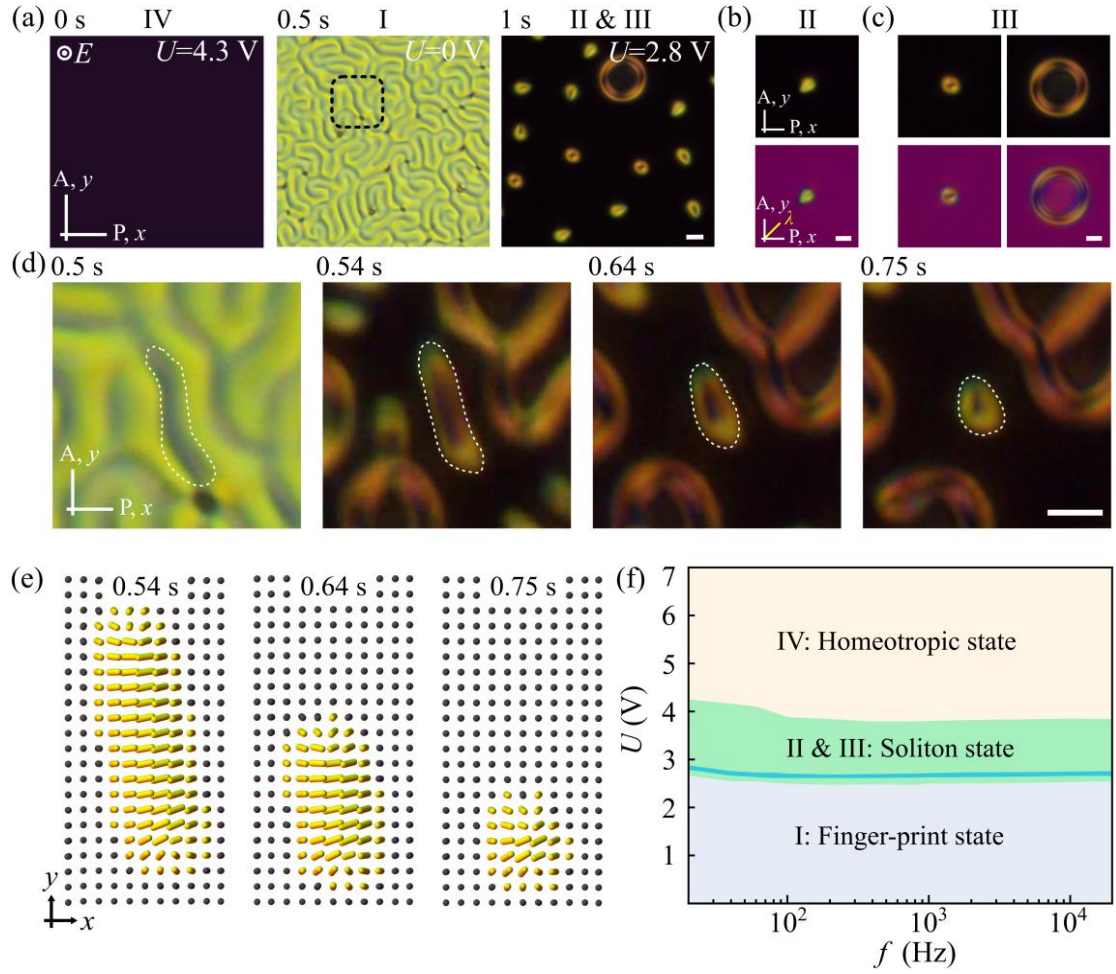

**Fig. 1. Generation of polar solitons. (a)** The system is first subjected to 4.3 V to achieve a homeotropic state. After turning off the field, the system relaxes into a TIC state within 0.5 s, followed by the formation of polar solitons at 2.8 V ($f = 20$ Hz, $T = 25°C$, $d = 5.5$ μm). "P" and "A" represent the polarizer and analyzer directions, respectively. **(b)** POM of a polar soliton and **(c)** skyrmionic particles. Images taken with a first-order red plate compensator (530 nm) show the slow axis ($\lambda$) represented by the yellow rod, oriented at 45° to the crossed polarizers ($U = 2.8$ V, $f = 20$ Hz, $T = 25°C$, $d = 5.5$ μm). The scale bar represents 5 μm. **(d)** Contraction of a polar soliton from a fingerprint texture, showing the details in the dashed section of state I in Fig. 1a, at 2.8 V. **(e)** Simulation of the generation of polar solitons from a fingerprint texture. **(f)** Phase diagram illustrating various states under different voltages. The gray region (I) corresponds to fingerprint states, the blue region (II) indicates where polar solitons are observed, the green region (III) includes skyrmionic particles and polar solitons, and the yellow region (IV) represents the homeotropic state.

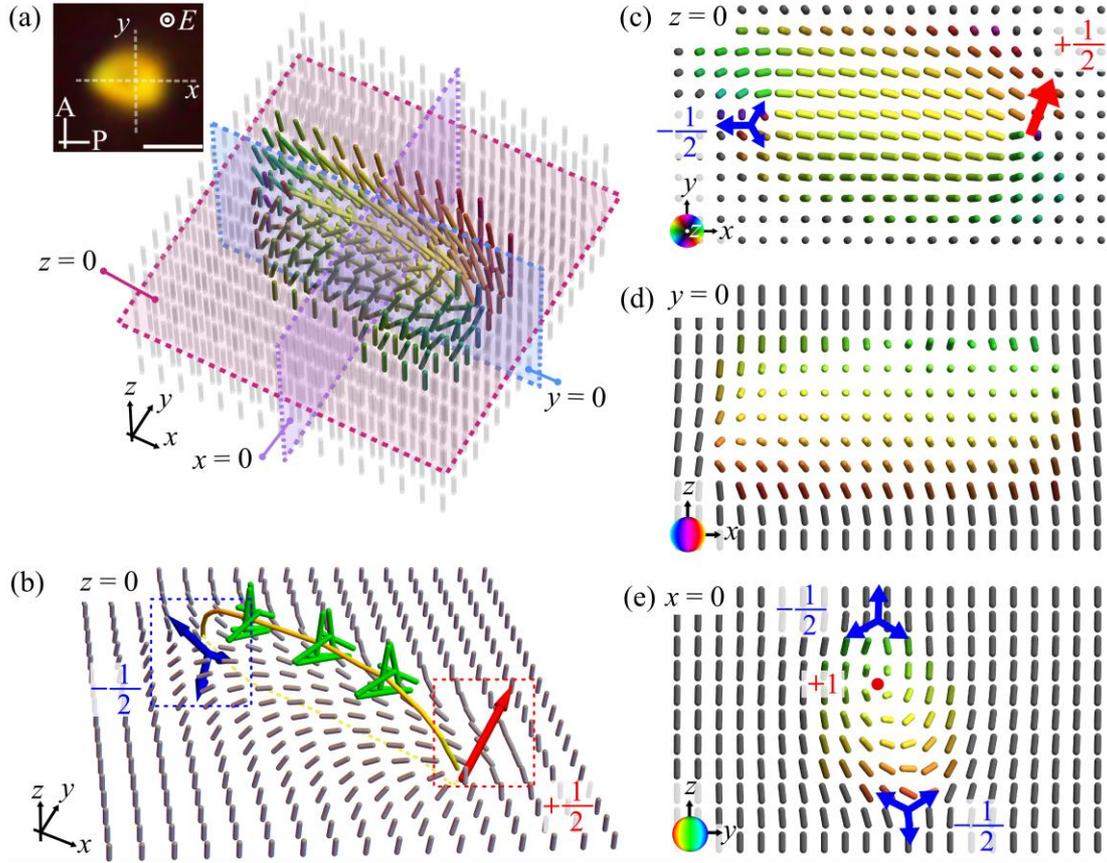

**Fig. 2. Director field of polar solitons. (a)** POM image and corresponding 3D structure of a single polar soliton, resembling a corn kernel ($U = 2.8$ V, $f = 20$ Hz, $T = 25°C$, $d = 5.5$ μm). Scale bar represents 5 μm. **(b)** Defect positions of the polar soliton: the blue triple arrow represents the $-1/2$ defect, the red arrow represents the $+1/2$ defect, and the yellow line indicates the defect line within the polar soliton. The green arrow shows the director field along this defect line. **(c)** Director field distribution of the soliton in the *xy*- plane ($z = 0$). **(d)** Director field distribution of the soliton in the *xz*- plane ($y = 0$). **(e)** Director field distribution of the soliton in the *yz*- plane ($x = 0$). The director field is color-coded according to orientations, as shown in the order parameter space (bottom left).

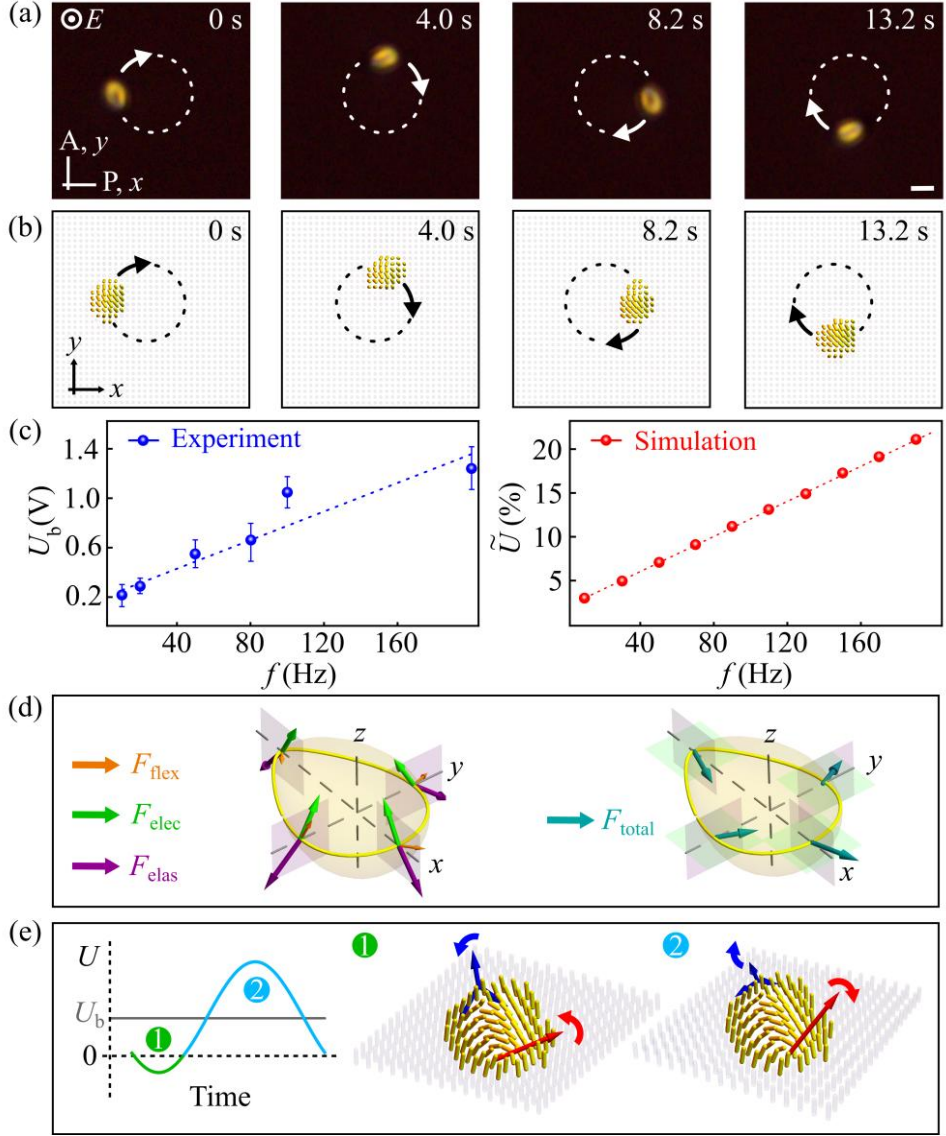

**Fig. 3. Circular motion of polar solitons. (a)** Experimental POM images and **(b)** simulation results showing the circular motion of polar solitons at different time points ($U = 2.8\,\text{V}$, $f = 20\,\text{Hz}$, $T = 25°C$, $d = 5.5\,\mu\text{m}$). **(c)** Relationship between the threshold bias voltage for initiating circular motion and the electric field frequency, including corresponding simulation data. **(d)** Analysis of the average forces over one period of the electric field at various positions along the polar soliton. **(e)** Evolution of the $+1/2$ and $-1/2$ defects within a single voltage cycle, highlighting structural changes in the polar soliton. Scale bar: 5 μm.

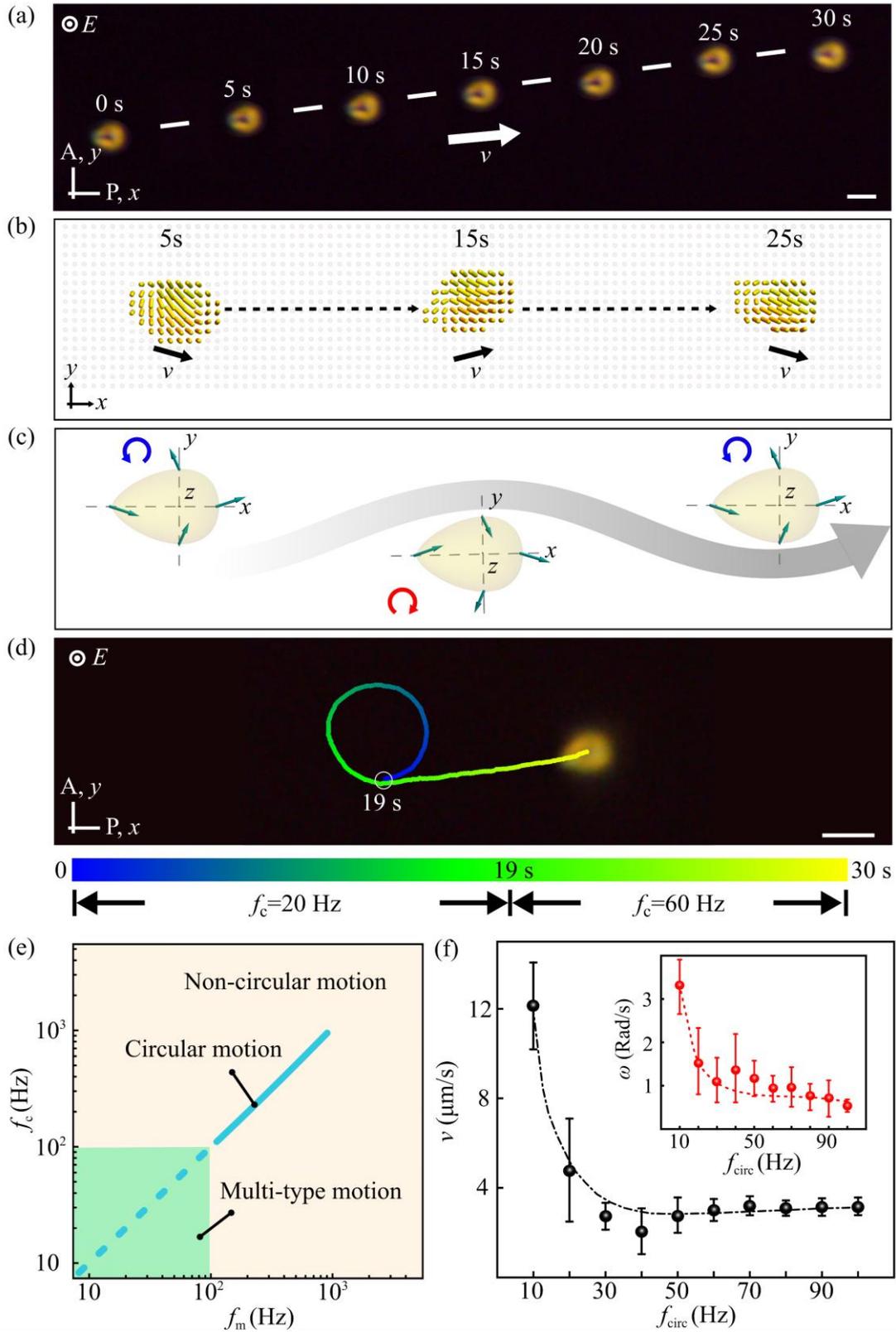

**Fig. 4. Modulation of polar soliton motion. (a)** Sequential snapshots of soliton linear motion at 5-second intervals ($U = 2.8$ V, $f_c = f_m = 20$ Hz, $T = 25°C$, $d = 5.5$ μm). **(b)** Simulation of soliton moving along a linear trajectory. **(c)** Force analysis of polar solitons during linear motion: solitons alternately perform clockwise and counterclockwise circular motions at short range, transitioning to linear motion at long range. **(d)** Transition from circular to linear motion induced by adjusting the carrier frequency of the modulated voltage ($U = 2.8$ V, $T = 25°C$, $d = 5.5$ μm). **(e)** Phase diagram of soliton motion modes: the green region indicates coexistence of circular and non-circular motion; the blue line represents stable circular motion when carrier and modulation frequencies match; the yellow region indicates the absence of circular motion. **(f)** Relationship between the linear and angular velocities of soliton circular motion and the modulation frequency. Scale bar: 5 μm.

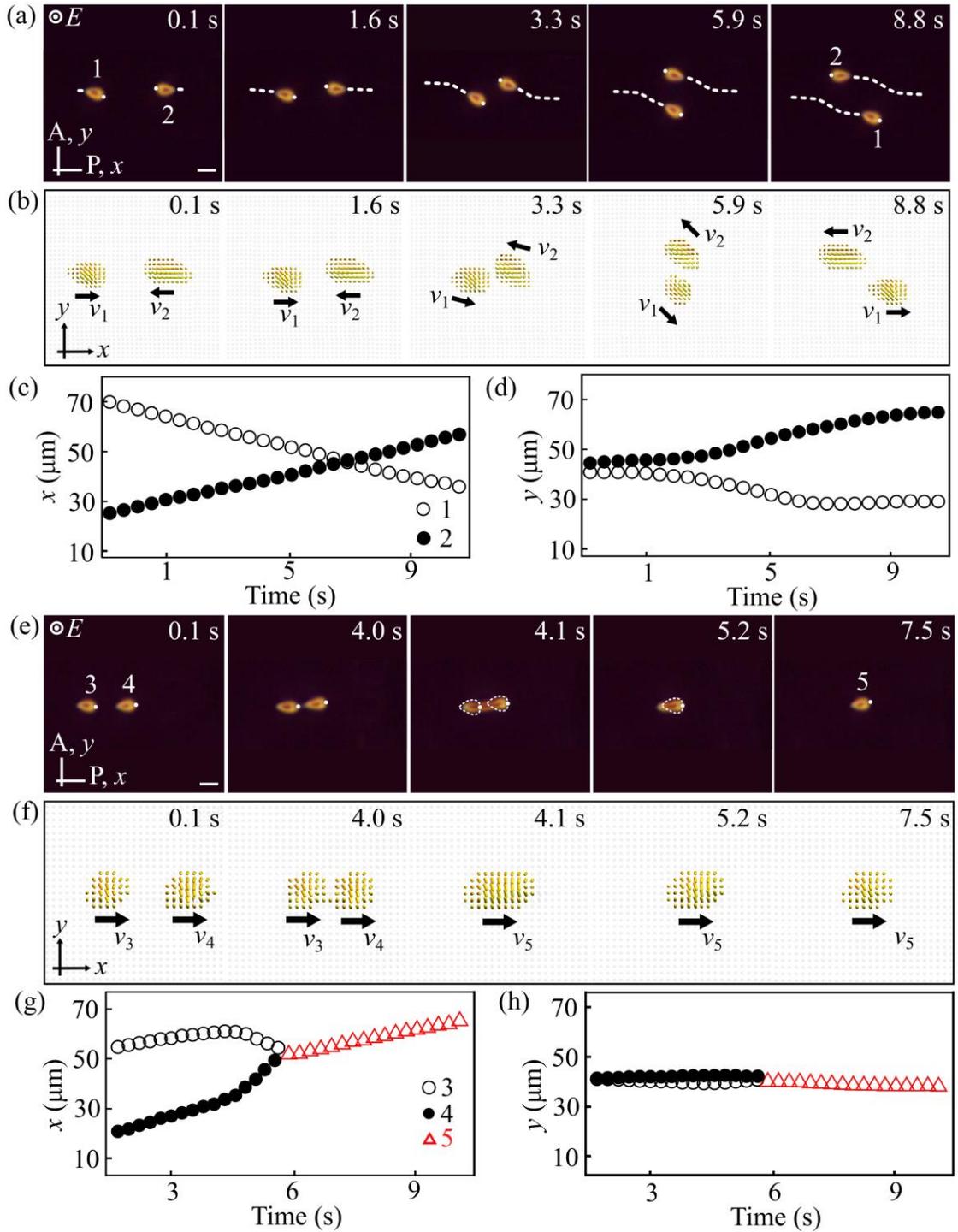

**Fig. 5. Interaction of polar solitons. (a)** POM images showing the collision of two oppositely moving polar solitons, labeled "1" and "2" ($U = 2.8$ V, $f_m = 20$ Hz, $f_c = 2$ kHz, $T = 25$°C, $d = 5.5$ μm). **(b)** Simulated director field of the two oppositely moving polar solitons during the collision. **(c, d)** Displacement of solitons "1" and "2" along the $x$- and $y$- axes over time. **(e)** POM images showing the attraction and fusion of two polar solitons, labeled "3" and "4", resulting in a new soliton labeled "5" ($U = 2.8$ V, $f_m = 20$ Hz, $f_c = 2$ kHz, $T = 25$°C, $d = 5.5$ μm). **(f)** Simulated director field of the attraction and fusion of two polar solitons and the resulting soliton. **(g, h)** Displacement of solitons "3", "4", and "5" along the *x*- and *y*- axes over time. Scale bar: 5 μm.

*Supplementary Information*

# Polar solitons in a nonpolar chiral soft matter system

Jia-Hao Chen, Xing-Zhou Tang, Yang Ding, Susanta Chakraborty,

Satoshi Aya, Bing-Xiang Li, Yan-Qing Lu

**Table of contents**

# #1. Supplementary Figures S1-S15

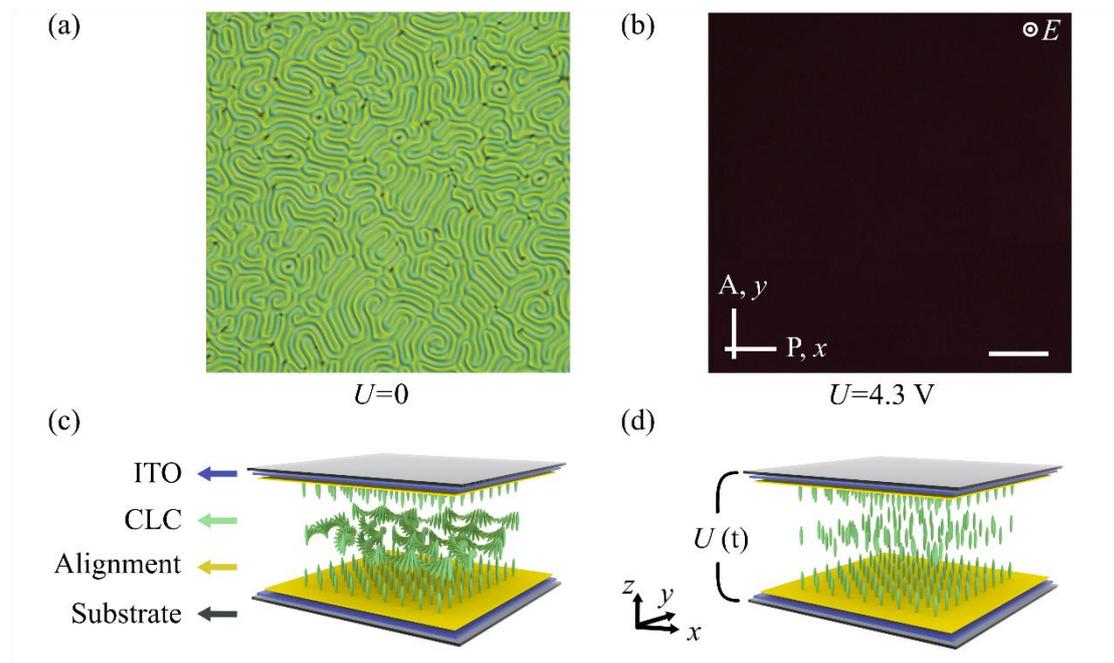

**Supplementary Fig. S1.** Experimental polarizing micrographs of **(a)** fingerprint state and **(b)** homeotropic state ($U = 4.5$ V, $f = 20$ Hz, $T = 25°C$, $d = 5.5$ μm). **(c)** Schematic of a sample without voltage applied. **(d)** Schematic of a sample with voltage applied across the LC using transparent electrodes on the inner surfaces of the confining substrates. Scale bar: 20 μm.

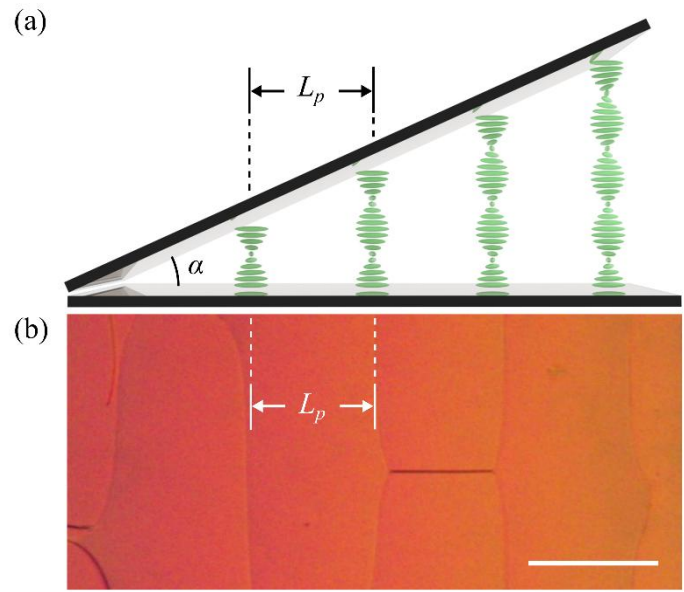

**Supplementary Fig. S2. (a)** Schematic illustration of the Grandjean-Cano cell. **(b)** Micrograph of a stripe-wedge Grandjean-Cano cell filled with chiral nematic liquid crystal, showing a gradual increase in cell thickness from left to right ($T = 25°C$). Scale bar: 80 μm.

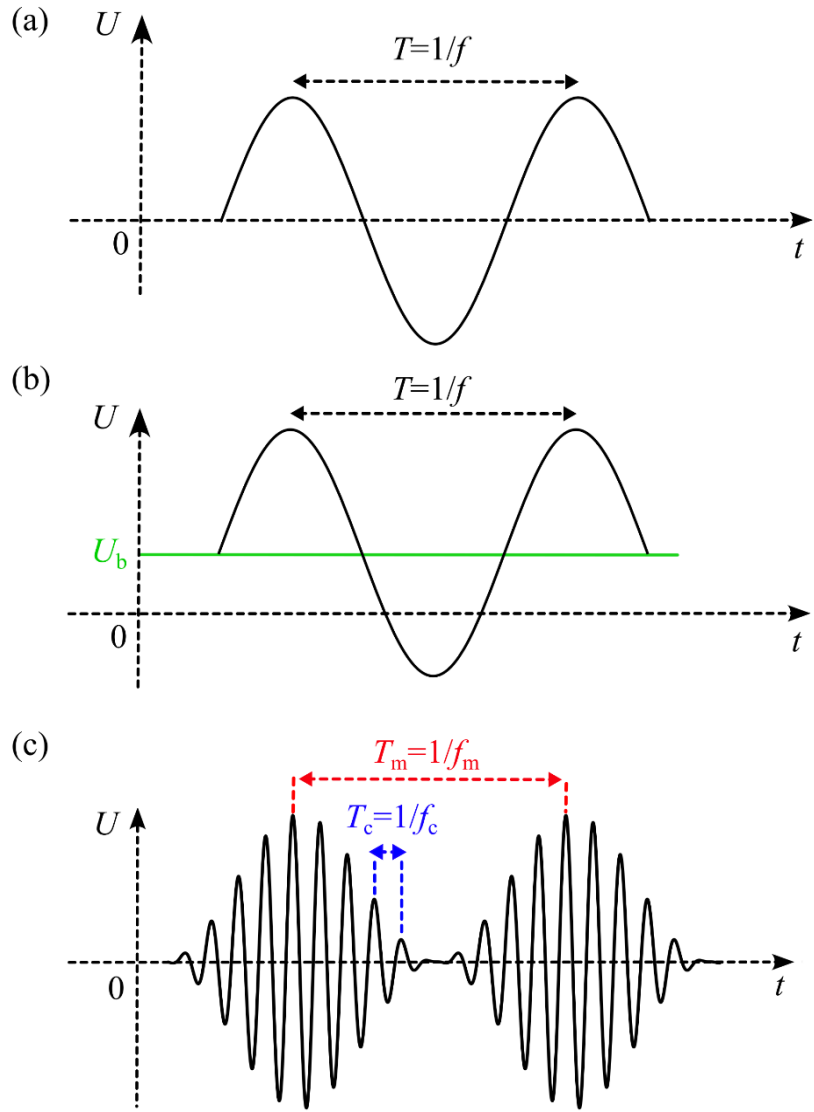

**Supplementary Fig. S3.** Voltage waveforms for AC electric fields. **(a)** Sine wave without bias voltage, under which polar solitons exhibit minimal movement. **(b)** Sine wave with a bias voltage, enabling circular motion of polar solitons. **(c)** Modulated waveform, supporting multiple types of soliton motions.

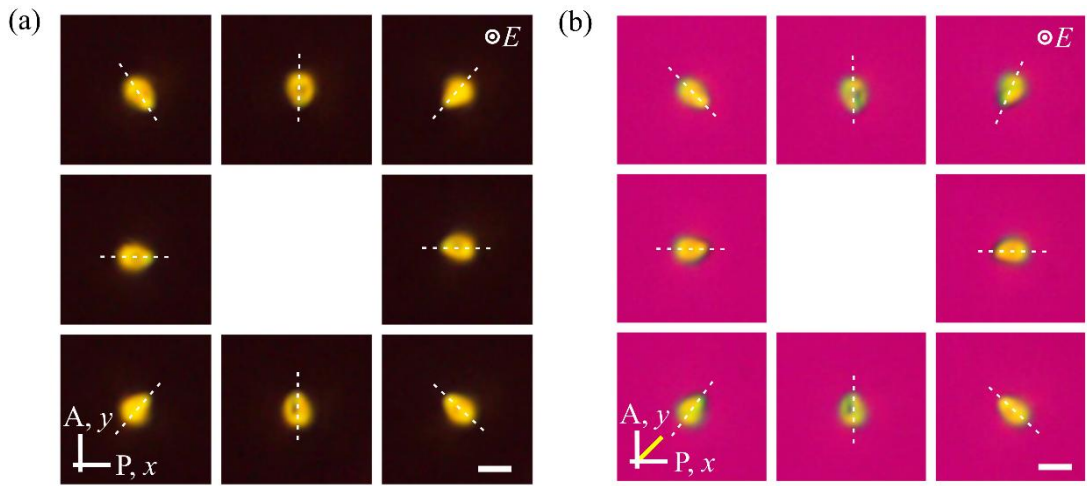

**Supplementary Fig. S4.** Polarizing optical micrographs of a single polar soliton. **(a)** Image under crossed polarizers. **(b)** Image with crossed polarizers and a first-order red plate compensator (530 nm), showing the direction of the slow axis (λ) indicated by the yellow rod, oriented at $45°$ to the crossed polarizers ($U = 2.8$ V, $f = 20$ Hz, $T = 25°C$, $d = 5.5$ μm). Scale bar 5 μm.

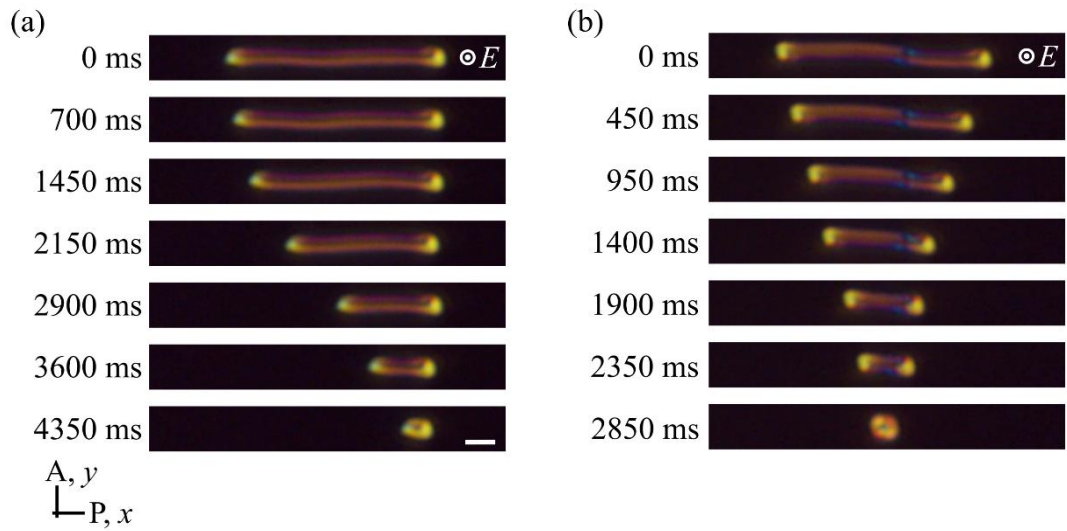

**Supplementary Fig. S5.** Contraction process of elongated fingerprint textures. **(a)** Contraction of a polar soliton. **(b)** Contraction of a skyrmionic particle ($U = 2.8$ V, $f = 20$ Hz, $T = 25°C$, $d = 5.5$ μm). Scale bar: 5 μm.

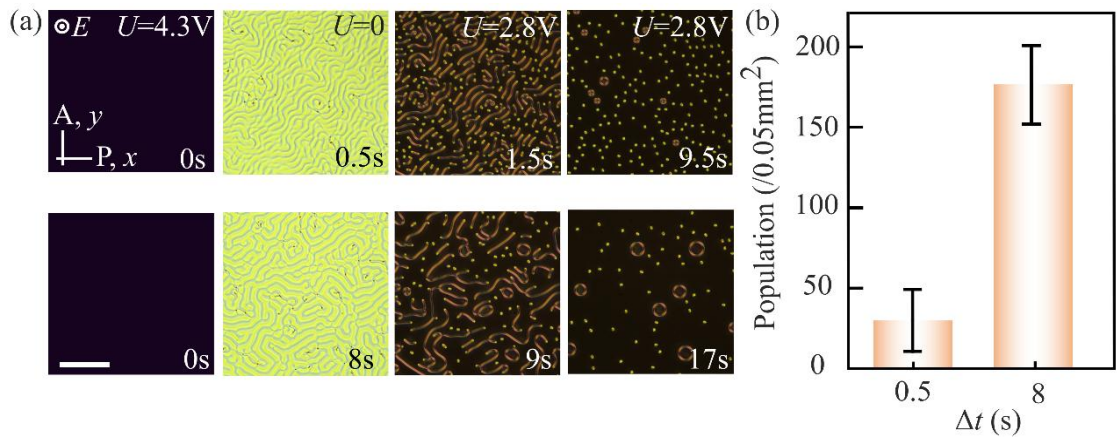

**Supplementary Fig. S6. (a)** Polar solitons generated under varying relaxation times. **(b)** Density of solitons as a function of relaxation time ($U = 2.8$ V, $f = 20$ Hz, $T = 25°C$, $d = 5.5$ μm). Scale bar: 20 μm.

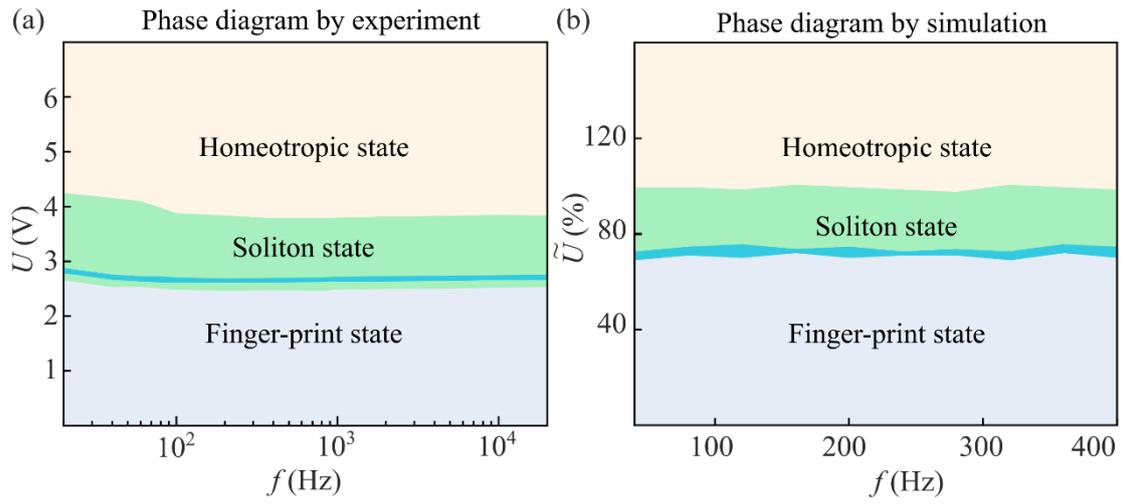

**Supplementary Fig. S7.** Phase diagram of various states under different voltages. **(a)** Experimentally observed states ($T = 25°C$, $d = 5.5$ μm). **(b)** Simulated phase diagram showing corresponding states.

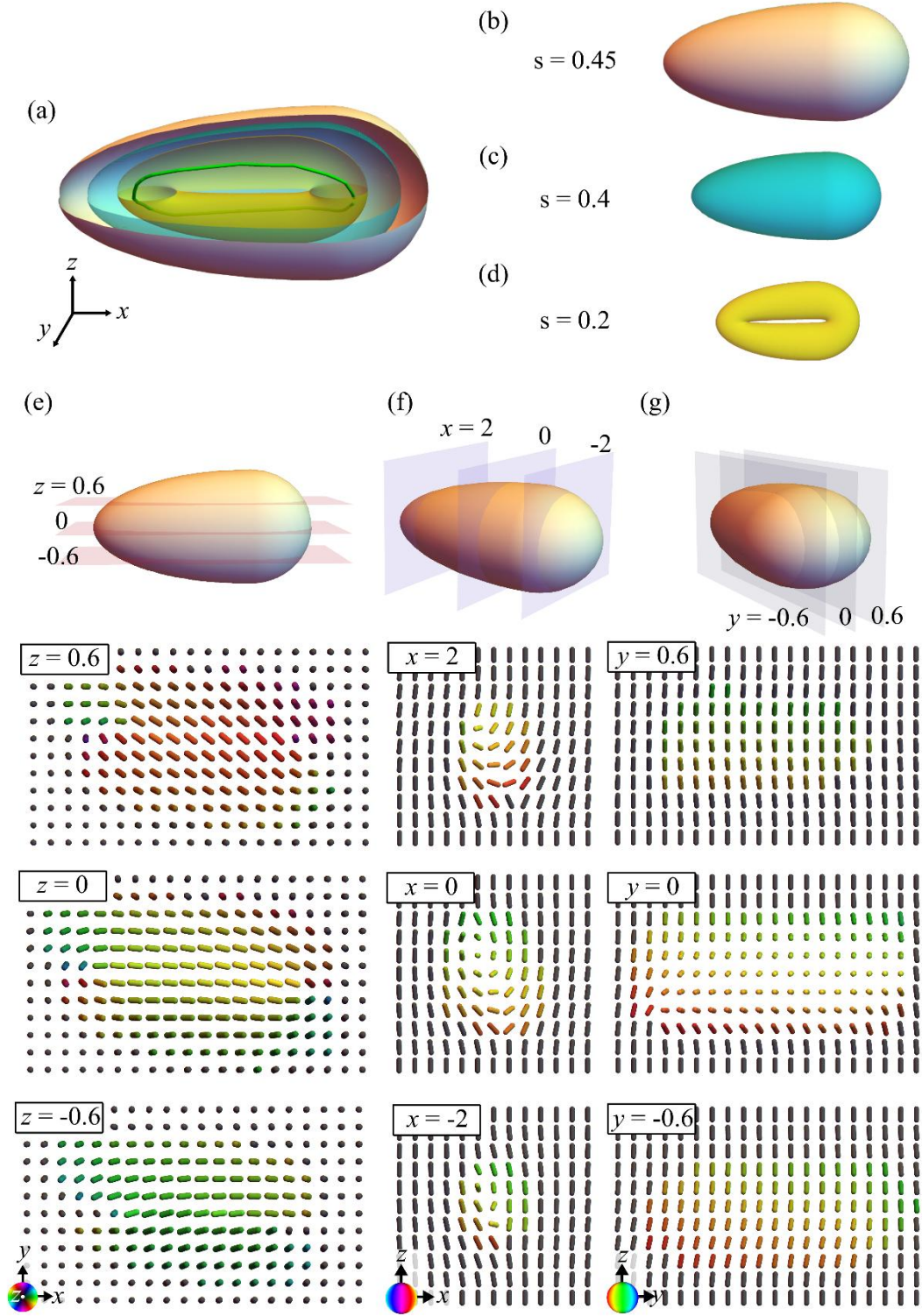

**Supplementary Fig. S8. (a)** A soliton shell model based on the scalar order parameter. From outside to inside **(b)** s = 0.45, **(c)** s = 0.4, **(d)** s = 0.2. The director field of polar solitons along **(e)** $z = -0.6, \ 0, \ 0.6$. **(f)** $x = -2, \ 0, \ 2$. **(g)** $y = -0.6, \ 0, \ 0.6$.

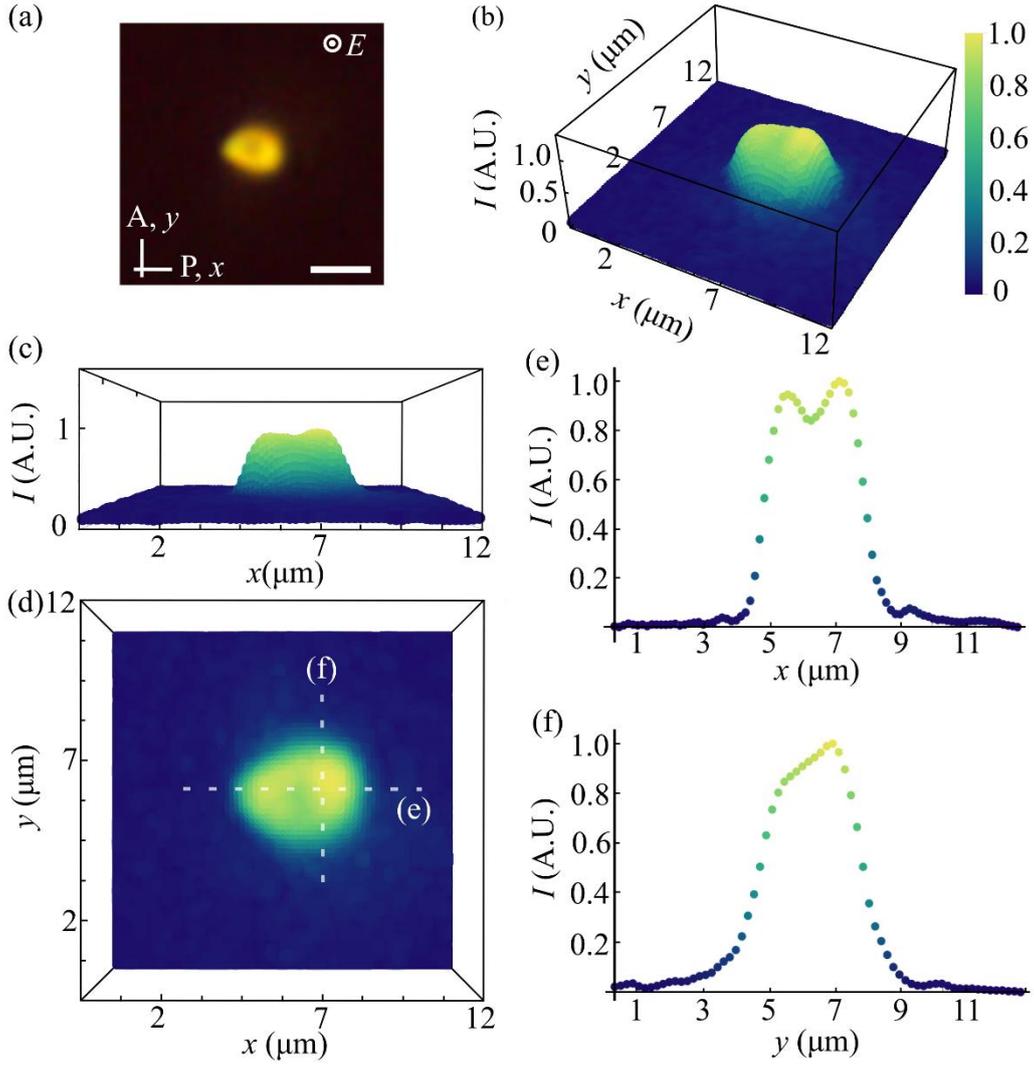

**Supplementary Fig. S9.** Analysis of the light intensity of a polar soliton. **(a)** Polarizing optical micrograph of a polar soliton under crossed polarizers ($U = 2.8$ V, $f = 20$ Hz, $T = 25°C$, $d = 5.5$ μm). Scale bar: 5 μm. **(b)** Isometric view, **(c)** side view, and **(d)** top view of the light intensity distribution of the soliton in **(a)**. **(e)** Line intensity distribution along the *x*- direction and **(f)** along the *y*- direction in **(d)**.

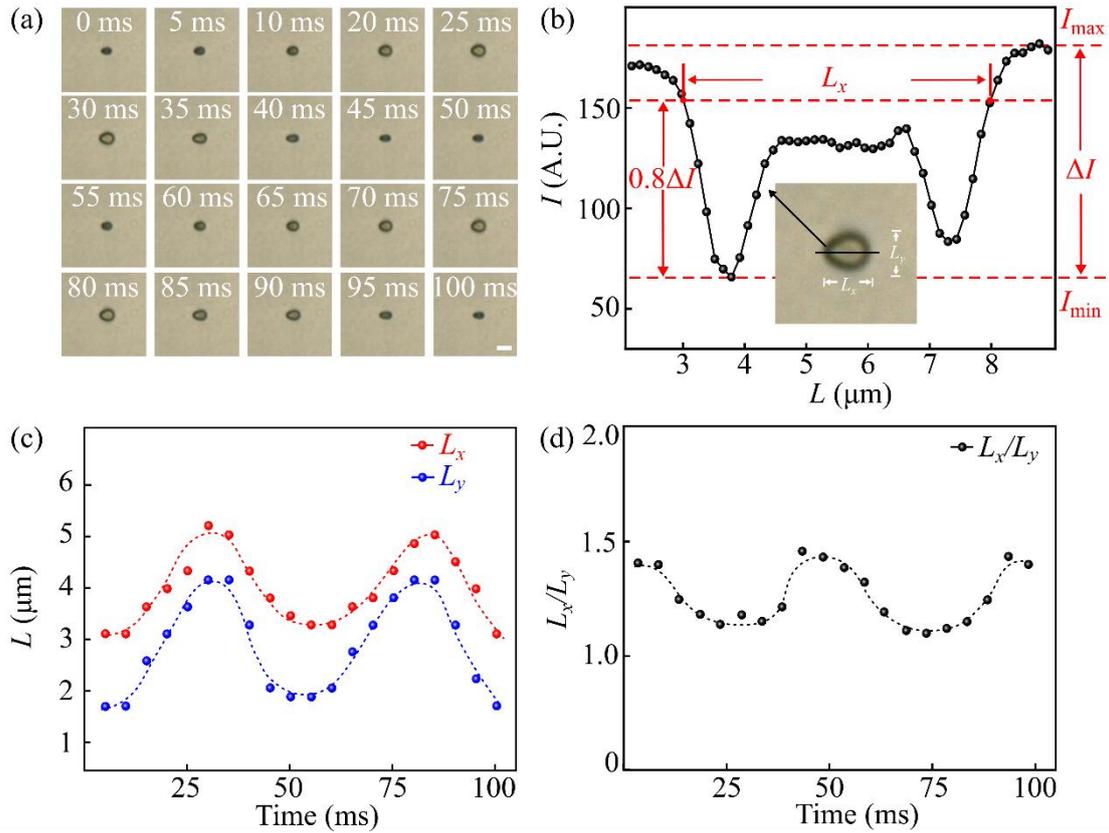

**Supplementary Fig. S10.** Changes in polar solitons within one voltage cycle. **(a)** Bright-field microscopy images (taken at the absence of polarizers) of a polar soliton at different time points ($U = 2.8$ V, $f = 10$ Hz, $T = 25°C$, $d = 5.5$ μm). Scale bar: 5 μm. **(b)** Definition of soliton size: Linear intensity distribution analysis is performed along the soliton's long or short axis. The maximum and minimum light intensities are denoted as $I_{max}$ and $I_{min}$, respectively, with the intensity difference defined as $\Delta I = I_{max} - I_{min}$. The soliton's size ($L_x$ or $L_y$) is defined as the distance between the two points where the intensity distribution function reaches $I_{min} + 0.8\Delta I$. **(c)** Variations of $L_x$ and $L_y$ during one voltage cycle. **(d)** Ratio of $L_x$ to $L_y$ over the cycle. Frame rate: 200 fps.

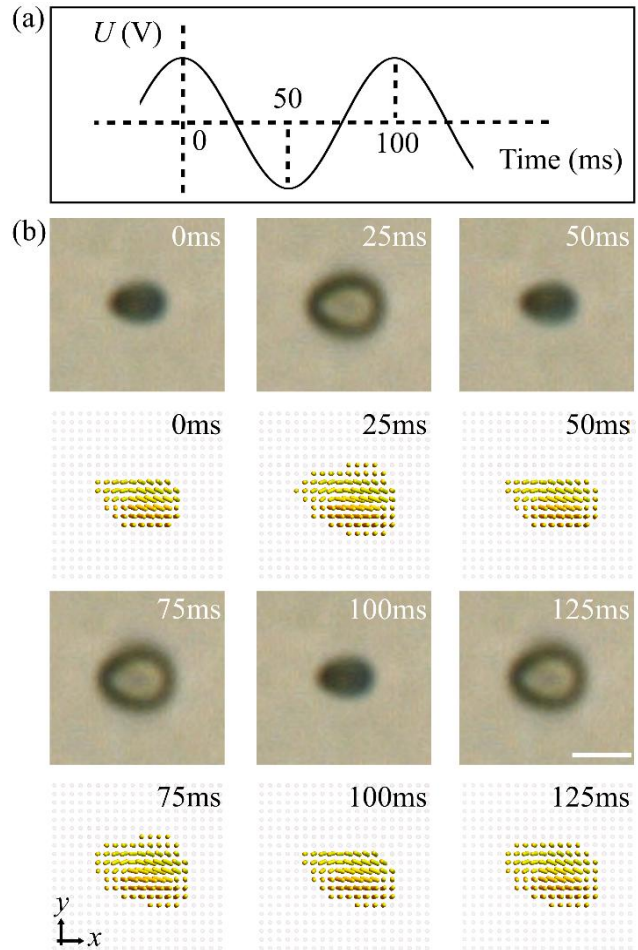

**Supplementary Fig. S11. (a)** The variation of the sinusoidal alternating current that generates polar solitons over time. **(b)** Numerically simulated director field and corresponding bright-field microscopy images of a polar soliton within one voltage cycle. ($U = 2.8$ V, $f = 10$ Hz, $T = 25°C$, $d = 5.5$ μm). Scale bar: 5 μm.

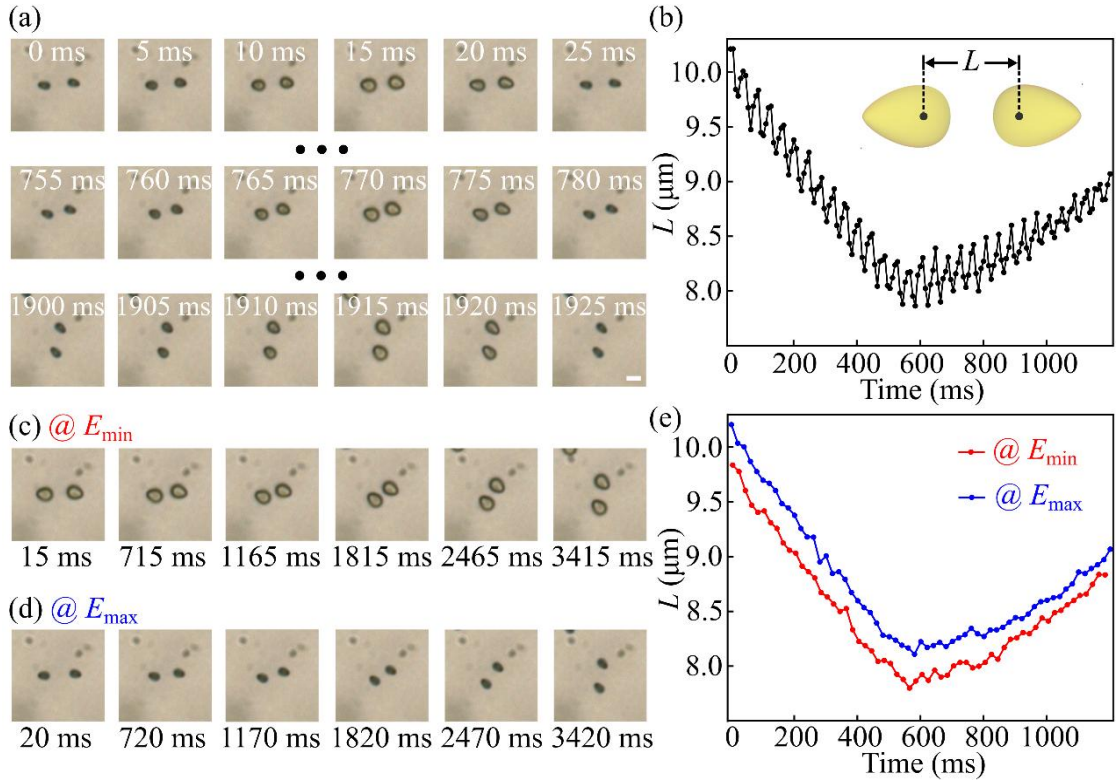

**Supplementary Fig. S12.** Collision dynamics of polar solitons. **(a)** Bright-field images showing the collision of a pair of polar solitons at different time points. **(b)** Distance between two solitons over time, defined as the distance between the intersection points of the soliton's long and short axes. **(c)** Collision process captured at the same frame of each voltage cycle when the solitons are largest. **(d)** Collision process captured at the same frame of each voltage cycle when the solitons are smallest ($U = 2.8$ V, $f_m = 20$ Hz, $f_c = 2$ kHz, $T = 25°C$, $d = 5.5$ μm). Scale bar: 5 μm. **(e)** Temporal evolution of the distance between the two solitons, corresponding to (c) and (d). Frame rate: 200 fps.

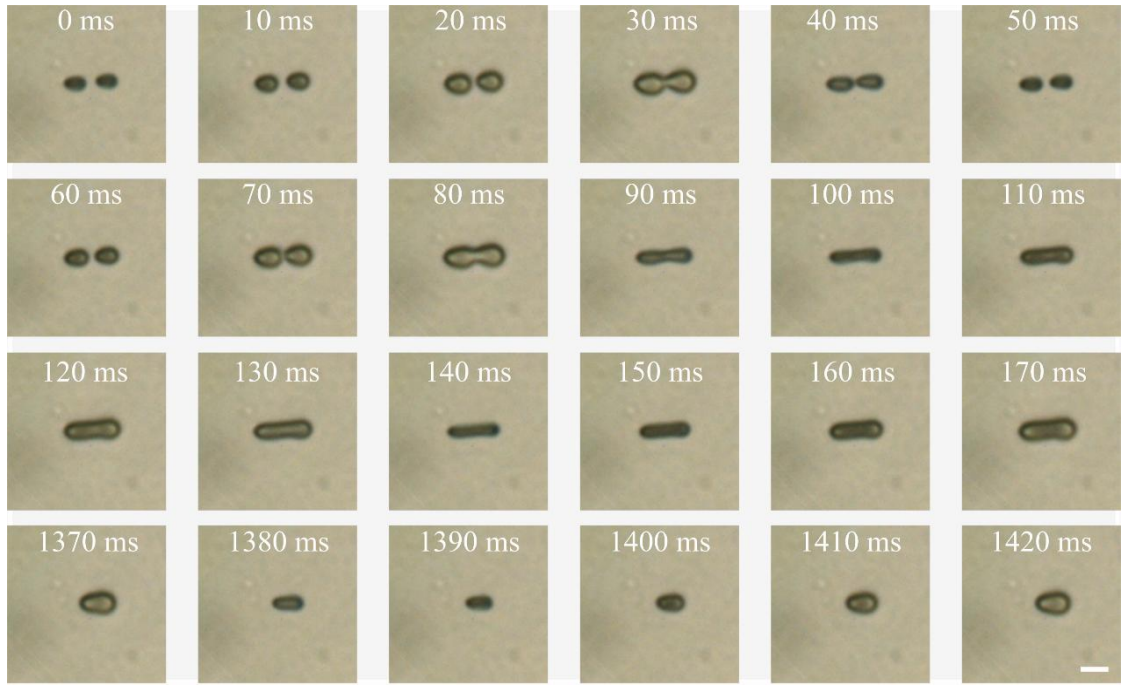

**Supplementary Fig. S13.** Attractive interaction of a pair of polar solitons in bright field. The frame rate is 200 fps. ($U = 2.8$ V, $f_m = 20$ Hz, $f_c = 2$ kHz, $T = 25°C$, $d = 5.5$ μm). Scale bar: 5 μm.

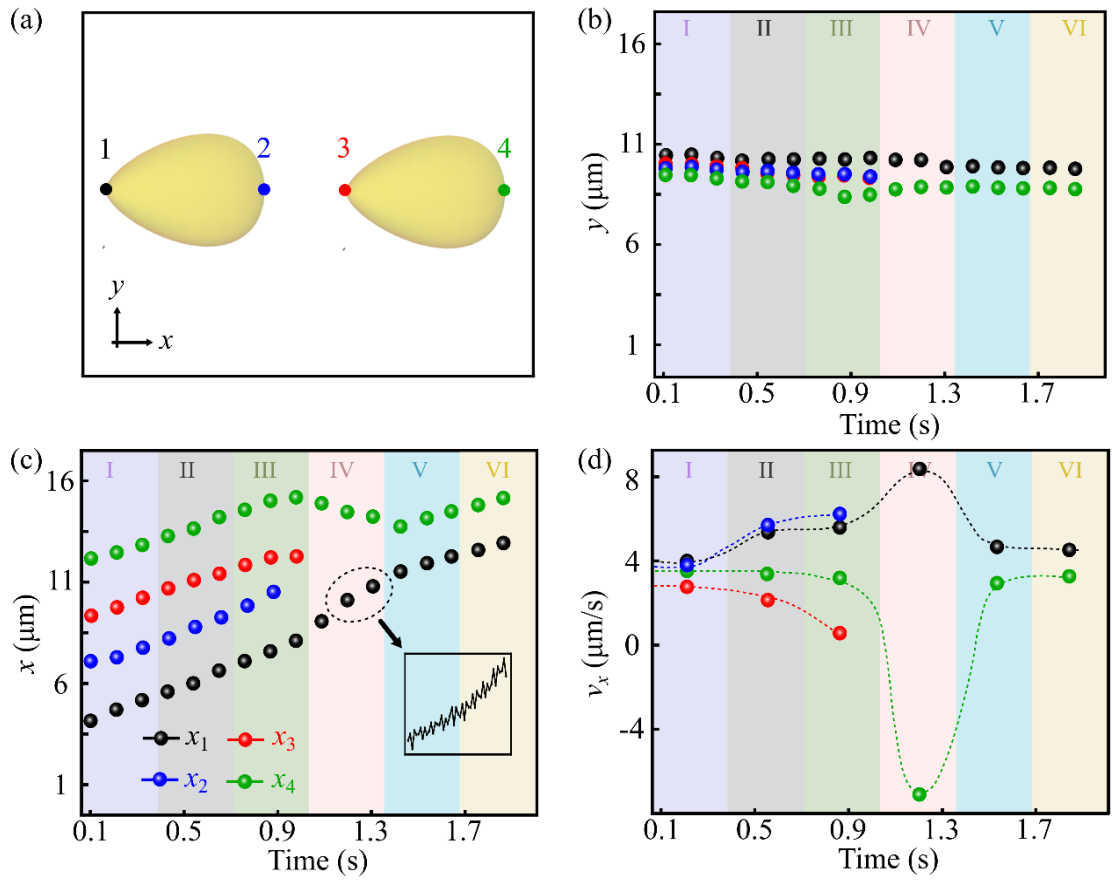

**Supplementary Fig. S14.** Temporal relationship between the head and tail positions of a pair of attracting polar solitons. **(a)** Four key points of the head and tail positions of a pair of attracting polar solitons. **(b)** Temporal changes in the *y*- coordinates of the four key points. **(c)** Temporal changes in the *x*- coordinates of the four key points and **(d)** corresponding velocity.

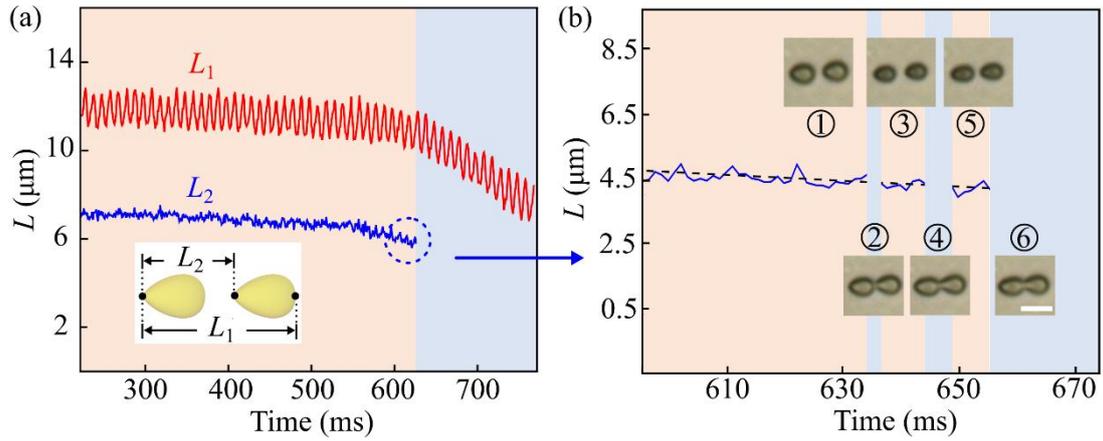

**Supplementary Fig. S15.** Temporal evolution of distances between different ends of polar solitons during attraction and merging. **(a)** Changes in two distances over time: $L_1$, the distance from the tail end of the trailing soliton to the head of the leading soliton; $L_2$, the distance from the tail ends of both solitons. The pink background represents the period before merging, while the blue background indicates the merged state. **(b)** Detailed changes in $L_2$ between 600 and 680 ms. Frame rate: 200 fps ($U = 2.8$ V, $f_m = 20$ Hz, $f_c = 2$ kHz, $T = 25°C$, $d = 5.5$ μm). Scale bar: 5 μm.

# #2. Supplementary Notes

**Supplementary Note 1: Preparation of CNLCs and LC cells**

The chiral nematic liquid crystals utilized in this study were prepared by a nematic liquid crystal, E7 (purchased from Jiangsu Hecheng Display Technology), S811 (S-(+)-2-Octyl 4-(4-hexyloxybenzoyloxy) benzoate, purchased from Nanjing Leyao).

To prepare the mixture, E7 and S811 were individually dissolved in dichloromethane (DCM) and subsequently mixed together. The mass fractions of E7 and S811 were $97\,\text{wt\%}$ and $3\,\text{wt\%}$, respectively. The resulting solution was subjected to $20$ minutes of thorough ultrasonic treatment in an ultrasonic cleaner at a temperature of $34°C$. Subsequently, the mixture was heated and dried on a hot plate set at $80°C$ for $1$ hour to allow for the evaporation of the DCM solvent. After the evaporation process, the prepared mixtures were injected into the cells at $80°C$ in their isotropic phase.

In order to observe the solitons, the cells were composed of two glass plates with transparent indium tin oxide (ITO) electrodes of low resistivity (ranging between $10$ and $50\,\Omega/\text{sq}$). The inner surfaces of the glass plates were coated with polyimide 4070 (Nanjing Ningcui) at $2700$ rpm for $30$ s, followed by a 5-min prebake at $120°C$ and a 60-min bake at $220°C$, and then unidirectionally rubbed to achieve a planar alignment of the chiral nematic liquid crystals, resulting in a helical pitch that is perpendicular to the glass substrates. To control the cell thickness within a range of $d = 4.0\text{-}6.0$ μm, silica spheres were dispersed in UV glue.

**Supplementary Note 2: Pitch measurement of chiral nematic liquid crystal**

For accurate determination of *p*, a Grandjean–Cano cell is applied. The thickness of the cell still varies continuously along the opening direction of the wedge but, importantly, discontinuously in the perpendicular direction, Fig. S2. The pitch can be determined through:

$$p = 2L_p \tan\alpha \tag{S1}$$

Where $L_p$ is the distance between the closed disclinations, $\alpha$ is the opening angle of the wedge-shaped liquid crystal cell.

**Supplementary Note 3: Free-energy calculation**

The system's free energy is characterized as $F = \int [f_{\text{LdG}} + f_{\text{elas}} + f_{\text{flex}} + f_{\text{diel}}]\, dV$. The enthalpic term, essential for describing phase transitions and interactions within the liquid crystal matrix, is specified as[1]:

$$f_{\text{LdG}} = \frac{A}{2}\left(1 - \frac{U}{3}\right)\text{Tr}(\mathbf{Q})^2 - \frac{AU}{3}\text{Tr}(\mathbf{Q}^3) + \frac{AU}{4}(\text{Tr}(\mathbf{Q}^2))^2 \qquad (S2)$$

where $A$ and $U$ are phenomenological parameters reflecting the material's properties[2,3]. The elasticity contribution, capturing the energy due to distortions in the director field, is given by[4,5]:

$$\begin{aligned} f_{\text{elas}} =\ & \frac{1}{27S^2}(K_{33} - K_{11} + 3K_{22})\partial_l Q_{jk}\partial_l Q_{jk} + \frac{2}{9S^2}(K_{11} - K_{22} - K_{24}) \\ & \times \partial_k Q_{jk}\partial_l Q_{jl}\partial_k Q_{jk}\partial_l Q_{jl} + \frac{2}{9S^2}K_{24}\partial_l Q_{jk}\partial_k Q_{jl} \\ & + \frac{2}{27S^3}(K_{33} - K_{11})Q_{jk}\partial_j Q_{lm}\partial_k Q_{lm} + \frac{4}{9S^2}q_0 K_{22}\varepsilon_{jkl}Q_{jm}\partial_l Q_{km} \end{aligned} \qquad (S3)$$

here, $K_{11}$, $K_{22}$, $K_{33}$, and $K_{24}$ are the elastic coefficients for splay, twist, bend and saddle-splay, $S$ is the scalar order parameter, $q_0 = 2\pi/p_0$ reflects the chirality of liquid crystal. $\varepsilon_{jkl}$ is the Levi-Civita symbol, and the convention of summing over repeated indices is used.

The dielectric energy contribution, reflecting the interaction between the electric field and the dielectric properties of the material, is defined as[6,7]:

$$f_{\text{diel}} = -\frac{1}{2}\varepsilon_0\varepsilon_a E_i Q_{ij} E_j \qquad (S4)$$

with $\varepsilon_0$ and $\varepsilon_a$ denoting the vacuum permittivity and the permittivity anisotropy.

The flexoelectric contribution, influenced by the interaction between nematic distortion and polarization under an external electric field, is articulated as[8-10]:

$$\begin{aligned} f_{\text{flex}} =\ & -\frac{4}{3}\chi_0 E_i\partial_j Q_{ij} - \frac{4}{3}\chi_+ E_i(Q_{ik}Q_{jk}) - \frac{1}{3}\chi_2 E_k\partial_k(Q_{ij}Q_{ij}) \\ & -\frac{4}{9}\chi_- E_i\partial_j(Q_{ik}\partial_j Q_{jk} - Q_{jk}\partial_j Q_{ik}) \end{aligned} \qquad (S5)$$

**Supplementary Note 4: Flexoelectricity effect**

The flexoelectric energy density, stemming from the coupling between nematic distortion and electric polarization, is articulated in the following expression[8,10,11]:

$$P_i = \frac{1}{2}(e_{11} + e_{33})s\partial_j(n_i n_j) + \frac{1}{2}(e_{11} - e_{33})s(n_i \partial_j n_j - n_j \partial_j n_i) + r_1 \partial_i s + r_2(2n_i n_j - \delta_{ij})\partial_j s \quad (S6)$$

where $e_{11}$ and $e_{33}$ represent the splay and bend flexoelectric coefficients, respectively, and $r_1$ and $r_2$ are related to order polarization. The terms $r_1$ and $r_2$ denote the polarization due to gradients in the scalar order parameter $s$, highlighting the complex interplay between nematic order and flexoelectric effects.

Linking the flexoelectric and order parameters, the following relationships are derived:

$$\begin{aligned} e_{11} &= 2\chi_0 + s\chi_+ + s\chi_- \\ e_{33} &= 2\chi_0 + s\chi_+ - s\chi_- \\ r_1 &= \frac{1}{3}\chi_0 + \frac{5}{3}s\chi_+ + s\chi_2 \\ r_2 &= \chi_0 + s\chi_+ \end{aligned} \quad (S7)$$

providing a nuanced understanding of how flexoelectric effects contribute to the overall behavior of nematic liquid crystals.

The non-uniformity of the order parameter $s$ within the system is considered, enabling a comprehensive analysis of flexoelectric phenomena. This approach captures the full spectrum of interactions and effects that govern the behavior of polar solitons in liquid crystal environments.

# #3. Description of Movies

**Supplementary Movie S1**

Description: POM video of the process of generating polar solitons involves increasing the voltage to $U = 5$ V, bringing the system to a homeotropic state. The electric field is then turned off to transition into the TIC state, followed by the rapid application of an electric field at a voltage of $U = 2.8$ V. This results in the generation of numerous polar solitons. ($f_m = 20$ Hz, $f_c = 2$ kHz, $T = 25°C$, $d = 5.5$ µm). Real time video rate.

**Supplementary Movie S2**

Description: POM video of the process of generating polar solitons and skyrmionic particles ($f_m = 20$ Hz, $f_c = 2$ kHz, $T = 25°C$, $d = 5.5$ µm). Real time video rate.

**Supplementary Movie S3**

Description: POM video of the contraction process of the polar soliton (the one below) and the skyrmion (the one above). ($U = 2.7$ V, $f_m = 20$ Hz, $f_c = 2$ kHz, $T = 25°C$, $d = 5.5$ µm). Real time video rate.

**Supplementary Movie S4**

Description: Computer-simulated the contraction process of the polar soliton from finger texture.

**Supplementary Movie S5**

Description: POM video of the process of a polar soliton undergoing circular motion. ($U = 2.8$ V, $f_m = 20$ Hz, $f_c = 20$ Hz, $T = 25°C$, $d = 5.5$ µm). Play at 5× speed.

**Supplementary Movie S6**

Description: Computer-simulated the circular motion of a polar soliton.

**Supplementary Movie S7**

Description: Microscopy video of the changes of a polar soliton within a voltage cycle, captured by a high-speed camera under bright-field conditions. ($U = 2.8$ V, $f_m = 20$ Hz, $f_c = 2$ kHz, $T = 25°C$, $d = 5.5$ µm). Play at 0.2× speed.

**Supplementary Movie S8**

Description: POM video of the process of a polar soliton undergoing linear motion. ($U = 2.8$ V, $f_m = 20$ Hz, $f_c = 60$ Hz, $T = 25°C$, $d = 5.5$ μm). Real time video rate. Play at 5× speed.

**Supplementary Movie S9**

Description: Computer-simulated linear motion of a polar soliton.

**Supplementary Movie S10**

Description: POM video of the motion of the polar soliton which can be changed from circular to linear by adjusting the carrier frequency from 20 Hz to 60 Hz. ($U = 2.8$ V, $f_m = 20$ Hz, $T = 25°C$, $d = 5.5$ μm). Play at 5× speed.

**Supplementary Movie S11**

Description: POM video of the repulsive motion of two polar solitons when the angle between their initial velocity directions is approximately 180°. ($U = 2.8$ V, $f_m = 20$ Hz, $f_c = 2$ kHz, $T = 25°C$, $d = 5.5$ μm). Real time video rate.

**Supplementary Movie S12**

Description: Microscopy video of the repulsion of two polar solitons captured by a high-speed camera under bright-field conditions. ($U = 2.8$ V, $f_m = 20$ Hz, $f_c = 2$ kHz, $T = 25°C$, $d = 5.5$ μm). Play at 0.2× speed.

**Supplementary Movie S13**

Description: POM video of the attraction of two polar solitons. ($U = 2.8$ V, $f_m = 20$ Hz, $f_c = 2$ kHz, $T = 25°C$, $d = 5.5$ μm). Real time video rate.

**Supplementary Movie S14**

Description: Microscopy video of the attraction of two polar solitons captured by a high-speed camera under bright-field conditions. ($U = 2.8$ V, $f_m = 20$ Hz, $f_c = 2$ kHz, $T = 25°C$, $d = 5.5$ μm). Play at 0.2× speed.

# #4. Supplementary Reference